\documentclass[prd,aps,twocolumn,showkeys,tightenlines,notitlepage,nofootinbib,preprintnumbers,letterpaper,superscriptaddress]{revtex4-1} 
\usepackage{epsfig}
\usepackage{amsfonts}
\usepackage{amsmath}
\usepackage{graphicx}
\usepackage{url}
\usepackage{color}
\usepackage{amssymb,amsmath,slashed,color,graphicx,bm}
\usepackage[T1]{fontenc}
\usepackage{relsize}
\usepackage{hyperref}
\hypersetup{colorlinks,citecolor= niceblue,linkcolor= niceblue, urlcolor=niceblue}
\definecolor{niceblue}{rgb}{0.1,0.2,0.6}
\definecolor{Violet}{rgb}{0.58, 0.4, 1}

\usepackage[normalem]{ulem}

\begin{document}
\title{Implications of a matter-antimatter mass asymmetry in Penning-trap experiments}

\author{Ting Cheng}
\email{ting.cheng@mpi-hd.mpg.de}
\affiliation{Max-Planck-Institut f\"ur Kernphysik, Saupfercheckweg 1, 69117 Heidelberg, Germany}

\author{Manfred Lindner}
\email{manfred.lindner@mpi-hd.mpg.de}
\affiliation{Max-Planck-Institut f\"ur Kernphysik, Saupfercheckweg 1, 69117 Heidelberg, Germany}

\author{Manibrata Sen}
\email{manibrata.sen@mpi-hd.mpg.de}
\affiliation{Max-Planck-Institut f\"ur Kernphysik, Saupfercheckweg 1, 69117 Heidelberg, Germany}

\begin{abstract}
The Standard Model (SM) of particle physics, being a local, unitary and Lorentz-invariant quantum field theory, remains symmetric under the combined action of Charge, Parity, and Time Reversal (CPT) symmetry. This automatically implies that fundamental properties of particles and antiparticles should be equal in magnitude. 
These fundamental tenets of the CPT principle have been put to stringent tests in recent Penning-trap experiments, where the matter-antimatter mass asymmetry has been measured. In light of these recent advances, we compare the bounds arising on CPT invariance from kaon systems with those from Penning-trap experiments. Using a simple yet powerful argument of mass decomposition of hadrons, we show that bounds on quark-antiquark mass differences from kaon oscillations are way beyond the reach of Penning-trap experiments. We lay out a roadmap to discuss possible reformulations of our understanding of the SM in the case of a discovery of CPT violation by these precision experiments.

\end{abstract}


\maketitle

\section{Introduction}
The Standard Model (SM) of Particle Physics is a local, Lorentz invariant, Hermitian quantum field theory (QFT).  As explored in a series of celebrated papers~\cite{Pauli:1940zz,Luders:1954zz,Jost:1957zz}, one of the fundamental tenets of such a local, Lorentz invariant theory is the conservation of CPT symmetry, that is, invariance under the combined operations of charge conjugation, parity inversion, and time reversal. 
The conservation of CPT guarantees that physical properties of matter and antimatter are related, for example, their masses should be identical, their charges, if any, should be equal and opposite. In fact, the requirement that a Hermitian QFT is causal automatically warrants the existence of antimatter which should have the exact same mass of the corresponding matter field. Therefore, a test of whether there exists a mass asymmetry between matter and antimatter, aptly dubbed here as the matter-antimatter mass asymmetry (MAMA), automatically translates to a test of the sacred principle of CPT invariance, and in turn, the foundations of the SM.

Theoretically, a number of motivations exist for CPT symmetry to be exact, relating the properties of matter and antimatter. 
 However, the baryon asymmetry of the Universe implies a matter dominated Universe. This indicates that some form of asymmetry between matter and antimatter must have been introduced through a new \emph{yet-unknown} mechanism in the early Universe. While models of successful baryogenesis usually follow a CPT symmetric approach, focusing on the Sakharov conditions~\cite{Sakharov:1967dj}, a baryon asymmetry could also arise in thermal equilibrium in the presence of CPT violation, and baryon number violation~\cite{Bertolami:1996cq}. Additionally, extensions of the SM to incorporate a quantum theory of gravity often induces CPT violation~\cite{Tsujikawa:2013fta}. Phenomenological motivations include the search for violation of Lorentz invariance (LI), or violation of locality (L), leading to CPT violation~\cite{Colladay:1996iz,Colladay:1998fq,Kostelecky:2000mm,Coleman:1997xq,Charlton:2020xxc,Chaichian:2011fc,Dolgov:2012cm,Chaichian:2012hy, Fujikawa:2016her}.

Experimental tests of CPT invariance can be twofold: testing the properties of particles and antiparticles directly, or probing the indirect impact of CPT violation on other processes. In the absence of a specific model, it is difficult to compare different experimental results at the same footing. Therefore, in this \emph{letter}, we aim to bridge different mass measuring experiments in a model independent manner, such that any new physics resulting in  CPT breaking can be investigated through a bottom-up approach.  
By treating the constraints from each experimental result using dimensionless parameters, the strongest constraint on CPT symmetry itself is currently from the kaon oscillation experiments.
The MAMA parameter is the difference between the two diagonal terms of the Hamiltonian of $(K^0,\bar{K}^0)$ in flavor space, and is tested using the Bell-Steinberger relation constructed under assumptions of unitarity~\cite{BARMIN1984293,Workman:2022ynf}. Note that neutral kaon-antikaon oscillations involve a process where \emph{strangeness} is violated by 2 units at one loop, through a box-diagram~(see \cite{Nierste:2009wg}, and references therein). In such a scenario, the test of MAMA could be more sensitive to violations of the principle of locality, the underlying process being a loop process. On the other hand, neutrino oscillation experiments also provide another interesting test of CPT conservation. Here, for a given neutrino energy, and baseline of the experiment, oscillation parameters (mass-squared difference, mixing angles) are fitted separately for the neutrino and antineutrino spectra~\cite{Colladay:1996iz,Colladay:1998fq,DeGouvea:2002xp,Bahcall:2002ia,Kostelecky:2003xn,Kostelecky:2003cr,Datta:2003dg,Minakata:2005jy,Kostelecky:2011gq,Ohlsson:2014cha,Super-Kamiokande:2014exs,Diaz:2016fqd,deGouvea:2017yvn,Barenboim:2017ewj}. In this case, MAMA is measured with respect to the dispersion relation of the propagating neutrino, and therefore, provides a more sensitive probe of Lorentz-invariance violation~\cite{Liberati:2013xla}. For a discussion regarding tests of non-locality using neutrino oscillation experiments, see~\cite{Antusch:2008zj} and references therein. In all these cases, tests of CPT conservation usually quote the results in terms of the mass difference between the particles and the antiparticles. It is important to emphasize that while the definition of the measured ``mass'' may be different from experiment to experiment, these systems can eventually be sensitive to multiple underlying principles behind the breaking of CPT: non-locality, and/or Lorentz invariance violation (LI-V). 
On the other hand, not all CPT breaking effects are covered in our analysis, since we have narrowed down the arbitrariness of the origin of CPT violation by considering only MAMA tests.

Significant progress has also been achieved by precision experiments in testing the tenets of the CPT principle. The ALPHA experiment at CERN uses trapped antihydrogen to study its charge-neutrality~\cite{ALPHA:2014nwo,ALPHA:2016klx}, the ratio of gravitational mass to inertial mass~\cite{Bertsche:2018avg}, as well as a measurement for the hyperfine splitting in neutral antihydrogen~\cite{ALPHA:2017fsh}. The experiment has, hence far, shown that the SM is consistent with CPT conservation. Similar tests have been performed using other species of antimatter such as antiprotons, which can be trapped for longer times in Penning-traps~\cite{Brown:1985rh}. The BASE collaboration~\cite{BASE:2022yvh} at CERN measures the charge-to-mass ratio ($q/m$) of the proton/antiproton by comparing the cyclotron frequency ($\nu_c$) of a single antiproton $\bar{p}$ to those of a single negatively charged hydrogen $H^-$(to avoid systemic uncertainties by having the charge different from the antiproton). This is done using the Brown-Gabrielse invariance theorem: $\nu_c^2=\nu_+^2+\nu_z^2+\nu_-^2$, where $\nu_+,\nu_z, \nu_-$ are three eigenfrequencies, namely, the modified cyclotron frequency, the axial frequency and the magnetron frequency, and $\nu_c = 1/(2\pi)(q/m)B_0$. 
The Penning-trap captures a $H^-$ and a $\bar{p}$ produced at CERN; then by measuring the three eigenfrequencies under a homogeneous magnetic field of $B_0= 1.945\,{\rm T}$, one can obtain the ratio of the proton to antiproton's inertial mass. Since the inertial mass is directly measured in this setup, measurement of the annual modulation of the MAMA would relate directly to tests of extension of the weak equivalence principle, such as having scalar-tensor theories of modified gravity \cite{Hughes:1990ay,Charlton:2020xxc}.

With technology advancing in leaps and bounds, the sensitivities of the Penning-trap experiments are expected to get even better with time.
In light of these facts, it is important to compare the bounds on CPT arising from kaon systems, and neutrino systems, with those from current and upcoming precision experiments. 
In this \emph{letter}, we present a simple yet crucial argument relying on the mass decomposition of hadrons using the energy momentum tensor in Quantum Chromodynamics (QCD), which allows a hadron mass to be separable into individual quark contributions, and those coming from gluons, kinetic terms, as well as anomaly terms. To zeroth order, this allows the MAMA in the hadron system to be written as the mass difference between quark and antiquarks. Using this parameterization,
the existing bounds from kaon systems translates to bounds on the mass difference between quarks and antiquarks, which are well beyond the sensitivity of Penning-trap experiments to such quark-antiquark mass differences. Similar outcomes with regards to the differences in gravitational forces exerted on matter and antimatter have been presented in~\cite{Caldwell:2019icl}.
This emphasizes that any discovery by these precision experiments would warrant a serious reformulation of our understanding of QFTs in order to be compatible with the results from the kaon systems. 

Our discussion is organized as follows. In the next section, we outline the basic framework of decomposition of a hadron mass in QCD, and how it relates to the constituent quark masses. Then, we apply it to different systems: protons, kaons, and discuss the bounds arising on MAMA. We also compare bounds on CPT arising from neutrino oscillations. We finally discuss the theoretical implications of a positive measurement of MAMA from the Penning-trap experiments, and/or other experiments, and conclude.

\section{Hadron mass decomposition in QCD} The masses of hadrons in QCD are governed primarily by the interaction amongst the constituent quarks and gluons. In fact, while hadron masses have been quite well measured experimentally, disentangling the mass contributions of the individual quarks from the quark-gluon strong interaction dynamics is still a matter of active research. The incredible complexity of the problem arises from the fact that at low energies, QCD is non-perturbative, and hence one needs to resort to lattice computations to get a clearer understanding.

However, it is possible to obtain a phenomenological decomposition of the mass of a hadron using the energy-momentum tensor $(T_{\mu \nu})$ in QCD~\cite{Ji:1994av, Yang:2014xsa}. The symmetric, conserved energy-momentum tensor in QCD can be formally written in Euclidean space as
\begin{equation}
 T_{\mu \nu}= \frac{1}{4}\bar{\psi}\gamma_{(\mu} \overleftrightarrow{D}_{\nu)}\psi + F_{\mu\alpha}F_{\nu\alpha}- \frac{1}{4} \delta_{\mu\nu}F^2,
\end{equation}
where $\overrightarrow{D}=\overrightarrow{\partial}_\mu+i g A_\mu$ and $\overleftarrow{D}=\overleftarrow{\partial}_\mu - i g A_\mu$, for a gluon field $A_\mu$ and coupling $g$, while the $()$ in the kinetic term denotes symmetrization over the indices. The QCD Hamiltonian operator can be defined through $T_{\mu \nu}$ as 
\begin{equation}
    H_{\rm QCD}=-\int d^3x\, T_{44}(x)~.
\end{equation}
This Hamiltonian can be decomposed into four contributions, coming from the kinetic energy of the quarks, the gluon field energy, the bare quark masses, and the QCD anomaly respectively, given by 
\begin{equation}
    H_{\rm QCD}= H_E + H_g + H_m + H_a~,
    \label{eq:Hdecomp}
\end{equation}
where 
\begin{eqnarray}
H_E &=&\sum_{q}\int d^3x\, \bar{\psi}_q(\overrightarrow{D}.\overrightarrow{\gamma})\psi_q ~, \label{eq:Hcomp1}\\
H_g &=&\int d^3x\, \frac{1}{2}(B^2-E^2) ~,\label{eq:Hcomp2}\\
H_m &=&\sum_{q}\int d^3x\, m_q \bar{\psi}_q \psi_q  ~,\label{eq:Hcomp3} \\
H_a &=&\, \int d^3x\,\left[\frac{\gamma_m}{4}\sum_{q}  m_q \bar{\psi}_q \psi_q - \frac{\beta(g)}{4g}(B^2+E^2)\right]. \,\,\, 
\label{eq:Hcomp4}
\end{eqnarray}
Here $\gamma_m $ is the anomalous mass dimension operator, and $\beta(g)$ is the QCD beta function. Using this decomposition, the masses of hadrons were estimated using lattice QCD for the first time to identify the contributions of different components~\cite{ Yang:2014xsa, Yang:2018nqn}.

The rest mass of a single hadron state $|x \rangle$ is defined as
\begin{equation} \label{Eq_Mx}
	m_x = 
	\frac{\langle x|H_{\rm QCD}|x\rangle }{\langle x| x \rangle}  ~.
\end{equation}
We stress that by focusing on such a mass measurement, we are narrowing down the arbitrariness of the origin of CPT violation to create a rather model independent bridge relating different mass testing experiments. In other words, $m_x$ does not include all CPT violating effects, for instance, terms that do not appear in the rest frame are not included.   
For example, a CPT transformation would transform a proton state ($|p\rangle$) onto an antiproton state ($|\bar{p}\rangle$), and vice versa. Hence, a non-zero mass difference between them would indicate CPT viloation. 
However, since the charge of the proton is decided solely by the valence quark charges, one can define a C$_q$PT transformation with the charge conjugation operator only acting quark fields and states (parity and time reversal operators still remain universal), which would do the same trick. Therefore, terms independent of quark fields in the mass decomposition formula, such as $H_g$, would only experience a PT transformation under C$_q$PT, which we take as invariant in the rest frame. This is the fundamental assumption that enters our calculation.

From the mass decomposition formula, it is clear that the masses of the quarks contribute to the hadron mass dominantly through the bare quark mass term in Eq.\,\ref{eq:Hcomp3}, and the anomaly term in Eq.\,\ref{eq:Hcomp4}, the remaining contributions being from the gluon field energy. The gluon energy contributions depend only on the gluon field, and does not depend dominantly on the quarks involved. As a result, when considering the mass difference between a hadron $x$, and its corresponding antihadron ${\bar{x}}$, the gluon contributions can be expected to cancel out at zeroth order. This mass difference should be dominated by $H_m$, followed by subdominant contributions from $H_a$ and $H_E$.
Note that the above formalism is derived under the assumption of a local and LI QFT; any violation from this might contribute to the quantum corrections in $H_a$. However, the arguments that follow will hold as these can be absorbed by a rescaling of the coefficients of $H_g$ and $H_m$. Namely, having the same field structure, we can combine the $\gamma$ and $\beta$ terms in $H_a$ with $H_m$ and $H_g$, respectively, and redefine the phenomenological coefficients. For instance, we can redefine $m_q\rightarrow m_q(1+\gamma_m)$, where $|\gamma_m| << 1$~\cite{Baikov:2014qja}.
Another equivalent way of stating the same is the following: a hadron mass can be assumed to be due to a combination of its bare quark masses, the gluon self-energy, and the quantum corrections received by these terms. The gluon self-energy term, as well as the quantum corrections are, to zeroth order, independent of the different constituent quarks of the hadron. As a result, when one takes the difference between the mass of a hadron and its corresponding antihadron, the difference due to the bare quark masses are the only terms to survive. Using the results of tests of CPT invariance, the hadron mass decomposition can immediately be used to put very stringent constraints on the mass difference between a quark and an antiquark. This is the underlying principle of our \emph{letter}. We elaborate on this in the following section.

\section{Comparison among the proton, the kaon and the neutrino sector}
 The BASE collaboration~\cite{BASE:2022yvh} measures 
the effective charge-to-mass ratio of protons and antiprotons to obtain limits on the mass asymmetry from their result $(q/m)_p/(q/m)_{\bar{p}} =-1.000000000003(16)$, 
\begin{equation} 
\label{Eq:protonB}
   \left|\frac{m_{\bar{p}}}{m_p}-1 \right| < 3\times 10^{-12}\,.
\end{equation}
Following our argument in the previous section, assuming the charges are equal and opposite for particles and antiparticles, this bound on mass difference can be translated onto the mass differences in the constituent valence quarks and antiquarks.
For estimation purpose, we consider three parameterizations of the MAMA in quarks. We define the difference in quark-antiquark masses as $\delta_q\equiv m_{\bar{q}}-m_q$, and the ratio between the quark-antiquark masses as $r_q \equiv m_{\bar{q}}/m_q$, where $q=s,d,u$ denote valence quarks. 
Here, the mass terms $m_q$ and $m_{\bar{q}}$ represent the pole energies of the propagator in the rest frame, which could be shifted differently for different spinor eigenstates. 
In particular, from Eq.~(\ref{Eq_Mx}), we can see that $m_q$ and $m_{\bar{q}}$ represent the energy eigenvalues for the corresponding spinor state of $x$ in the rest frame.

We can also parameterize the CPT violating term through, $\alpha$, where $\alpha$ is defined as  $m_x = m_0 (1+\alpha)$ for particles and $m_{\bar{x}} = m_0 (1-\alpha)$, for antiparticles. In that case,  
\begin{equation} 
\label{Eq:alpha}
	\alpha \equiv \left|\frac{m_{\bar{x}}-m_x}{m_{\bar{x}}+m_x}\right| 
	\simeq \left|\frac{\sum_j\delta_j}{2 m_x}\right|\,.
\end{equation}
The limits of the Penning-trap system can thus be translated into $| 2\delta_u+\delta_d|$ and $r-1$ assuming $r=r_d=r_u$. The left-hand-side of Eq.~\ref{Eq:protonB} can therefore be rewritten as $\left|\sum_{q} C_q \delta_q/m_p\right| $ or $\left|(r-1)\sum_{q} C_q m_q/m_p\right|$, where $C_q=\langle P|\bar{\psi}_q \psi_q |P\rangle = \partial m_p/\partial m_q$, and would be the same for proton and antiproton. Furthermore, since the sea quarks would also be identical between proton and antiproton, CPT breaking should be from the valance quarks with contributions proportional to the respective bare quark charges, namely, $C_u:C_d = 2:1$. The absolute value of $C_q$ can be estimated by considering an analogy with the mass splitting between proton and neutron ($\delta m^{\rm QCD}_{\rm N}$) induced by chiral symmetry breaking. Namely, the isovector scalar charge $g^{u-d}_s\sim 1$ \cite{FlavourLatticeAveragingGroupFLAG:2021npn}  for  $g^{u-d}_s =\delta m^{\rm QCD}_{\rm N}/ \delta m_{ud}$, where $\delta m_{ud}$ denotes the up quark-down quark mass splitting.
On the other hand, $C_u$ and $C_d$ can also be estimated through the pion-nucleon $\sigma$-term, $\sigma_{\pi N}=(m_d+m_u)/2 \langle N| \bar{\psi_u}\psi_u + \bar{\psi_d} \psi_d| N \rangle$, from pion-nucleon scattering experiments or through lattice calculations, see \cite{FlavourLatticeAveragingGroupFLAG:2021npn,Gupta:2021ahb} for a review.  
With the purpose of order-of-magnitude comparison between different systems, Table \ref{tab:MAMA} is listed conservatively to a range such that inputs from the chiral symmetry breaking analogy, the $\sigma$-term and $\langle H_m \rangle/m_p$ from the mass decomposition \cite{Yang:2018nqn}, can all be included.
Nonetheless, $\alpha$ is directly related to the experimental result, hence it would be independent of non-perturbative estimations, and is set as $\alpha<1.5 \times 10^{-12}$ by Eq.~\ref{Eq:protonB}.

\begin{table}
\begin{tabular}{ |c|c|c|c| } 
 \hline
 MAMA & Proton & Kaon & Neutrino \\
 \hline
 $|\sum_j\delta_j|$ (MeV) & $\mathcal{O}(10^{-10}-10^{-9})$ & $\mathcal{O}(10^{-16})$ & $\mathcal{O}(10^{-9})$ \\ 
 \hline
  $\delta$ (MeV) & $\mathcal{O}(10^{-10}-10^{-9})$ & trivial & $\mathcal{O}(10^{-9})$ \\ 
 \hline
 $r-1$& $\mathcal{O}(10^{-11}-10^{-10})$ & $\mathcal{O}(10^{-18})$ &$\mathcal{O}(10^{-1})$ \\
 \hline
 $\alpha$ & $\mathcal{O}(10^{-12})$ &  $\mathcal{O}(10^{-19})$  & $\mathcal{O}(10^{-2})$ \\
 \hline
\end{tabular}
 \caption{Theoretical limits on the different parameters of CPT violation. $|\sum_j\delta_j|$ denotes the sensitivity of MAMA for each experiment, which is  $ |2\delta_u+\delta_d|$, $|\delta_s-\delta_d|$ and $(|\delta_2|,|\delta_3|)$ from the Penning-trap, neutral kaon oscillation and neutrino oscillation experiments, respectively. $\delta$ assumes all $\delta_j$s are identical as such value. The other quantities are as defined in the text.}
 \label{tab:MAMA}
 \end{table}

The neutral kaon oscillation system offers a stronger limit on such MAMA. The Bell-Steinberger relation~\cite{Bell:1966vvu} allows a direct connection between the Hamiltoninan eigenstates, and the decay amplitudes in the $K^0-\bar{K}^0$ system.
From this, the limits on the mass difference between the kaon and the antikaon can be expressed using the diagonal elements of the Hamiltonian as~\cite{Workman:2022ynf}
\begin{equation}
\label{Eq:KaonB}
   |m_{K^0}-m_{\bar{K}^0}|<4 \times 10^{-16}\, {\rm MeV}\,.
\end{equation}
Following the same line of reasoning as that for the proton system, the entries for the kaon system in Table \ref{tab:MAMA} can be estimated for only bare quark contributions ($C=1$), or all valence quark contribution ($C\sim 2.1$ considering \cite{Aguilar:2019teb}) or even including sea quarks ($C\sim 2.5$ considering \cite{Aguilar:2019teb,Yang:2014xsa}).
Here, we assume $C=C_s=C_d$.
Note that Eq.~\ref{Eq:KaonB} would be trivially satisfied when $\delta_s=\delta_d$.
A rather engineered way to escape stringent bounds set by the neutral kaon oscillation experiments is to have an additional symmetry which sets $\delta_s$ (nearly) identical to $\delta_d$.
On the other hand, $\alpha$ would be independent of $C_q$ and would be limited to $\alpha<4.0 \times 10^{-19}$.
Clearly, the bounds arising from neutral kaon oscillation systems are roughly seven orders of magnitude stronger than that of the Penning-trap experiment. This is due to a combination of the lower mass of the kaon, as well as the stronger experimental bounds existing on this system from tests of CPT conservation. 
Note that the above results should not be considered as precise calculations, but more as estimates.
What is important to note, however, is that the naive bounds from kaon systems are \emph{orders of magnitude} above proton systems, and hence clearly set a much more stringent bound on MAMA than what can be possibly achieved by such Penning-trap experiments in the foreseeable future.

While the discussion has been centered around hadronic systems so far, it is also worthwhile to consider the impact of CPT invariance on MAMA for the neutrino sector. The main advantage in considering neutrinos is that they are fundamental particles, and hence are not plagued by the uncertainties existing in QCD. Furthermore, neutrino masses are sensitive to new physics at higher mass scales, where it is possible to imagine that the underlying theory might be non-local and/or LI-V. 
Neutrino oscillation experiments can test CPT invariance by its effect on the mass-squared differences~\cite{Colladay:1996iz,Colladay:1998fq,DeGouvea:2002xp,Bahcall:2002ia,Kostelecky:2003xn,Kostelecky:2003cr,Datta:2003dg,Minakata:2005jy,Kostelecky:2011gq,Ohlsson:2014cha,Super-Kamiokande:2014exs,Diaz:2016fqd,deGouvea:2017yvn,Barenboim:2017ewj}. A dedicated analysis, using a combination of solar neutrino data, and KamLAND reactor antineutrino data gives~\cite{Barenboim:2017ewj,Tortola:2020ncu}
\begin{equation}
\label{Eq:NeutrinoSolB}
    \Delta m^2_{21} - \Delta \bar{m}^2_{21} < 4.7\times 10^{-5}\,{\rm eV}^2\,,
\end{equation}
while long-baseline, and short-baseline experiments can set
\begin{equation}
\label{Eq:NeutrinoAtmB}
    \Delta m^2_{31} - \Delta \bar{m}^2_{31} < 2.5\times 10^{-4}\,{\rm eV}^2\,.
\end{equation}
For a given value of the lightest neutrino mass from zero to the upper limit from cosmological bounds, this translates to bounds for MAMA parameters of the two heavier neutrinos shown in Table~\ref{tab:MAMA}. The order-of-magnitude estimations hold for both neutrinos.

Note that in \cite{Barenboim:2017ewj}, the bounds arising on CPT violation from Eq.\,(\ref{Eq:NeutrinoSolB}-\ref{Eq:NeutrinoAtmB}) were compared to the quantity $m^2_{K^0}-m^2_{\bar{K}^0}$, which is the quantity in the Lagrangian that determines the oscillation in the kaon systems~\cite{Murayama:2003zw}. As a result, the authors of \cite{Barenboim:2017ewj} obtained that bounds on CPT violation in the neutrino sector are stronger than the kaon sector. On the other hand, we have used the parameterizations in Table~\ref{tab:MAMA}, which rely on the mass-difference and not the mass-difference-squared, as a fundamental parameter that CPT violation tests. As a result, we obtain that the bounds on CPT from neutrino system , while stringent, are not competitive to the kaon system to constrain MAMA. Of course, it is crucial to remember that these two bounds are fundamentally different: while the kaon is a composite particle in QCD, and the limits quoted in Eq.\,\ref{Eq:KaonB} offer a probe of QCD dynamics, the neutrinos are elementary fermions and hence can offer a clearer test of the fundamental symmetry.

\section{Theoretical Implications of CPT violation} 
\begin{figure}[!t]
    \includegraphics[width=0.5\textwidth]{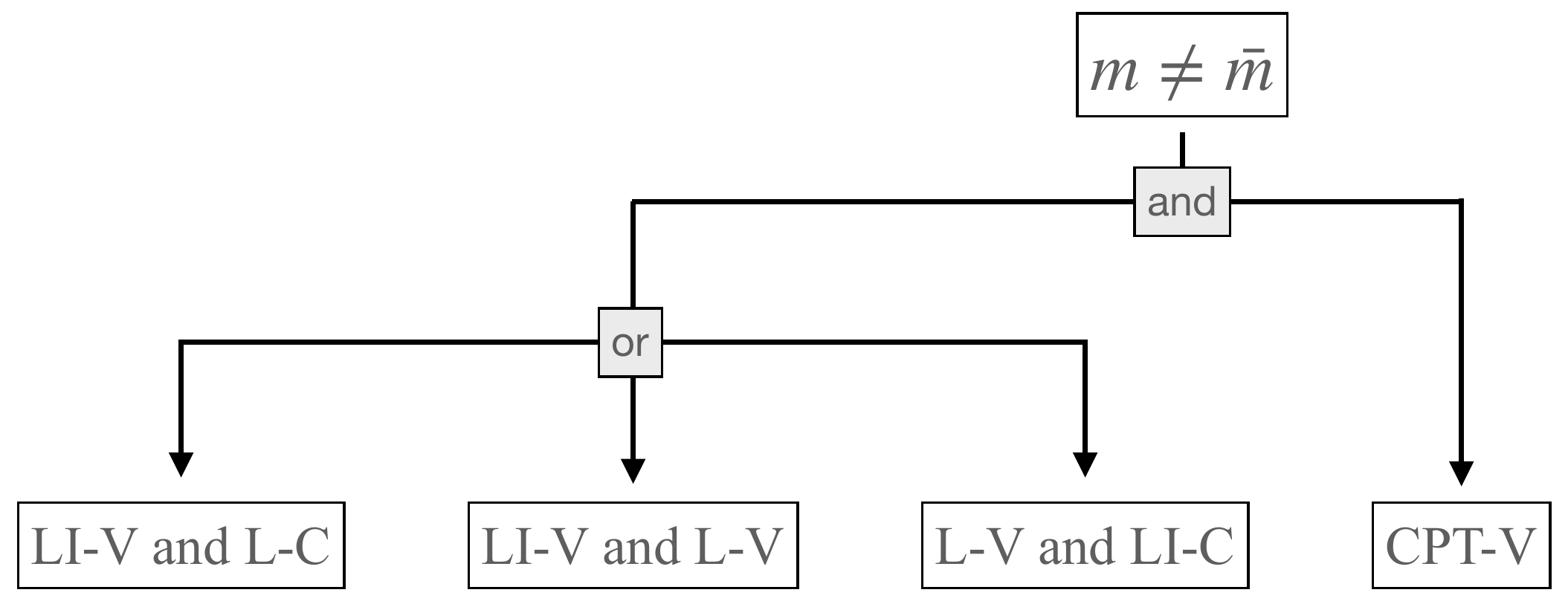}
    \caption{Flow chart for implications of a non-trivial asymmetry between the mass of matter and antimatter ($m\neq \bar{m}$). LI/L/CPT-V(C) means that Lorentz invariance/locality/CPT symmetry is violated (conserved). }
    \label{fig:flowchart}
\end{figure}

This brings us to an important question. 
In the SM and in any extension based on QFT, as long as quarks and neutrinos share the same source of CPT violation, we do not expect to see a signal of MAMA from either the Penning-trap experiment nor from neutrino oscillation due to the upper bound set by kaon oscillation. However, if the current Penning-trap experiments would observe a mass difference between protons and antiprotons, or in general, if a nontrivial MAMA is observed in any system, what kind of new physics could this signify? 
Immediately, this would indicate that CPT symmetry is broken, as shown in Fig.~\ref{fig:flowchart}, most likely by some new physics at high scale.
The CPT theorem states that in any \emph{local, Lorentz invariant} field theory, CPT is an exact symmetry. Clearly, if any of the above conditions are violated, CPT need not be an exact symmetry. In other words, unlike CP violation which is treated independently in different sectors, CPT violation stems from the fundamental principles not being exact, which is, in general, universal among different sectors. For instance, if Lorentz invariance fails in the quark sector, there is no obvious reason as to why it should not fail as well in the lepton sector.
Furthermore, as shown previously, one can connect MAMA in the hadronic system to that of the fundamental particles under the premise that a subset of the CPT transformation, $C_qPT$, is sufficient to reflect the hadronic state onto its antihadronic state. 
In this case, since only the rest mass is under consideration, the model independent mass decomposition formalism indicate that other terms would either be cancelled out or absorbed into $m_q/m_{\bar{q}}$ for the MAMA of hadrons. 
Here, we also don't consider spin dependence of the MAMA since we work in the rest frame, nonetheless, one can in principle do a spin dependent measurement to check this assumption.

\begin{figure}[!t]
    \includegraphics[width=0.35\textwidth]{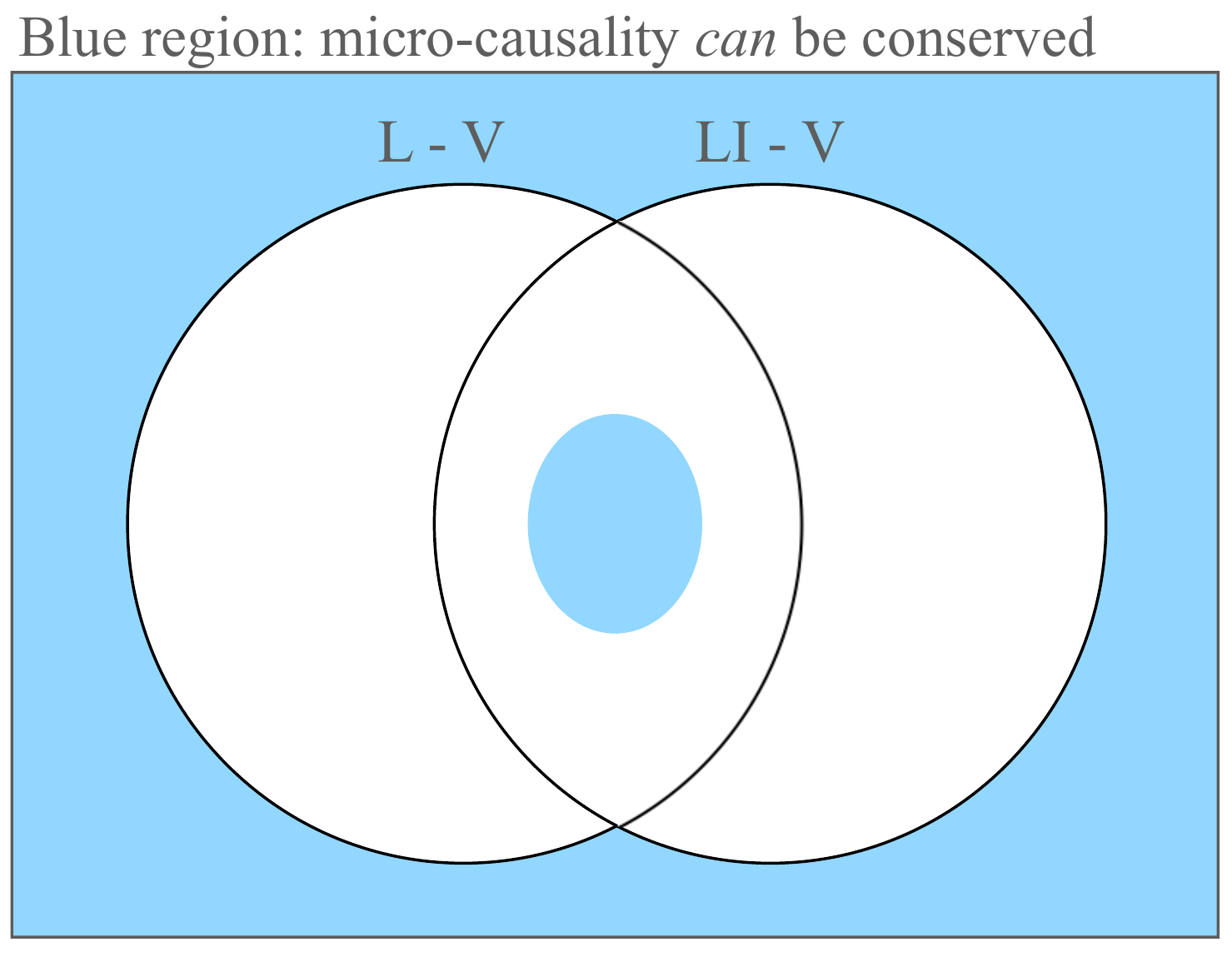}
    \caption{Both Lorentz invariant violation (LI-V) and locality violation (L-V) will break micro-causality, however, having both might cancel out such breaking~\cite{Kostelecky:2000mm, Tomboulis:2015gfa}}
    \label{fig:Venndiagram}
\end{figure}

In this section, we outline some of the fascinating new directions which can be explored in this context. This is intended to serve as a scaffolding for the establishment of new physics. Our arguments can be summarized in Fig.\,\ref{fig:flowchart}. 
A discovery of $m\neq\bar{m}$ would certainly imply violation of Lorentz invariance and/or locality violation. This necessitates an extensions of the SM, incorporating one or both of these tenets. 
Note that such extensions usually violate micro-causality, and hence must be embedded in a UV complete theory. However, for theories from spontaneous LI violation~\cite{Kostelecky:2000mm} or non-local field theories~\cite{Tomboulis:2015gfa}, causality would still be stably conserved at low energies, since the violation of micro-causality is confined in the high energy region.
This might be a way out in case of a positive discovery without restricting LI-V and L-V contributions for the breaking of micro-causality to cancel out, as shown in Fig.~\ref{fig:Venndiagram}.

The Standard Model Extension (SME)~\cite{Colladay:1996iz,Colladay:1998fq} aims to do this by writing down a phenomenological set of Lorentz-invariance violating operators.  Lorentz invariance can be broken either \emph{spontaneously}~\cite{Kostelecky:2000mm}, where some fundamental field in a Lorentz invariant theory takes a vacuum expectation value (vev), and breaks the symmetry, or \emph{explicitly}~\cite{Coleman:1997xq}, where the Lagrangian contains operators which break this symmetry from the beginning~\cite{Charlton:2020xxc}. In many cases, note that these vanilla scenarios predict the same charge-to-mass ratio for particles and antiparticles; a violation of this will make the theory non-causal~\cite{Greenberg:2002uu}. 
However, what matters is whether the low energy observables are causal, without going into the details of the high energy theory.  

To demonstrate how one can link a certain model to our MAMA parameter, we take the minimum SME (mSME) as an example. The CPT symmetry breaking terms in the mSME Lagrangian with a $H_m$ field structure, namely $a_\mu$ and $b_\mu$, satisfy the Dirac equation \cite{Colladay:1996iz}
\begin{equation} \label{Eq:SME_Dirac}
	(i\gamma_\mu D_\mu - m - a_\mu \gamma_\mu - b_\mu \gamma_5 \gamma_\mu)~\psi_q = 0.
\end{equation} 
The field operator can be written as 
\begin{equation}
	\psi_q(x)=\int \frac{d^3 \bold{p}}{(2\pi)^3}\sum_{s}\left(\frac{c^s_{\bold{p}}u_q^s(\bold{p})e^{-\mathrm{i}p_u^sx}}{\sqrt{2E_u^s}} +\frac{d^{s\dagger}_{\bold{p}}v_q^s(\bold{p})e^{\mathrm{i}p_v^sx}}{\sqrt{2E_v^s}} \right),\\
\end{equation}
where $c_{\bold{p}}^s,d_{\bold{p}}^{s\dagger}$ denote corresponding creation and annihilation operators, while $u^s_q, v^s_q$ are the spinors. The four-momenta are $p_u^s=(E_u^s,\bold{p})$, $p_v^s=(E_v^s,\bold{p})$, with $E_u^s$ and $E_v^s$ depending on $a_\mu$ and $b_\mu$ to satisfy the Dirac equation. For simplicity, we assume only $a_\mu$ is non-zero, then $E_u^s=E_u=\sqrt{m^2+(\bold{p}-\bold{a})^2}+a_0$ and $E_v^s=E_v=\sqrt{m^2+(\bold{p}+\bold{a})^2}-a_0$.
Hence, in the rest frame, $m_q = E_u(\bold{p}=0)$, $m_{\bar{q}} = E_v(\bold{p}=0)$ and the MAMA parameter $\delta_q=2\,(a_q)_0$.
From Table~\ref{tab:MAMA}, we get $(a_s)_0-(a_d)_0<2\times 10^{-16}$ MeV from kaon oscillation and $2(a_u)_0+(a_d)_0<2\times 10^{-10}$ MeV from the Penning trap bounds after taking all $C_q$s as one.
The energy eigenvalues for the case where both $a_u$ and $b_u$ are non-zero are given in \cite{Colladay:1996iz}, which would in general be spin dependent, giving four eigenvalues corresponding to the four spinors. However, in the rest frame, two of them would be degenerate with the other two, leaving the MAMA parameter still well defined.

The other alternative is to extend the Lagrangian such that it becomes a \emph{non-local} theory.
For e.g., \cite{Chaichian:2011fc,Dolgov:2012cm,Chaichian:2012hy, Fujikawa:2016her} considered a Lorentz invariant extension of QED, which however, has non-local operators, thereby leading to CPT-violation. These kind of theories predict different masses for particles and antiparticles at scales much below the non-locality scales, and hence can have direct implications for Penning-trap like experiments. However, these theories also suffer from non-unitarity~\cite{Chaichian:2012ga}, as well as micro-causality problems, and should be viewed strictly as an effective theory. 
Other possibilities which might lead to such mass difference can arise in composite quark theories~\cite{Larin:2020hih}, where different quarks pick up different CPT phases, or in models of extra dimension like string theory~\cite{Kostelecky:1988zi, Kostelecky:1989jw}, or in theories of quantum gravity~\cite{PhysRevD.21.2742}.

\section{Conclusion}

The conservation of CPT symmetry is a sacrosanct feature of any local, Lorentz invariant theory. Experiments are underway to test this theory by probing the charge-to-mass ratio of protons-antiprotons. The BASE collaboration at CERN has recently measured the proton/antiproton charge-to-mass ratio to an accuracy of parts per trillion. 
The collaboration plans to improve their sensitivity further by upgrading the stability of the  magnetic field, as well as using transportable antiproton traps. In this regard, it is crucial to examine the bounds arising on similar measurements from other experiments.

The most stringent bound on the extent of CPT violation comes from kaon-antikaon oscillation experiments. Using a simple phenomenological modelling of hadron masses, we have shown that the kaon bounds translate to bounds on the CPT violation parameter that are orders of magnitude beyond the limits from Penning-trap experiments. These bounds even supersede those from neutrino oscillation experiments. As a result, any positive results from Penning-trap experiments would definitely be a smoking-gun signal for fundamental new physics. This can either point towards a possible violation of CPT, thereby leading to a reformulation of the basic tenets of QFT, or a possible lack of understanding in the QCD sector. Additionally, if the positive result appears in the neutrino sector, it would point toward theories where the underlying source of CPT violation is different for neutrinos than for quarks. We have laid a tentative road-map for the future in the event of such a discovery.   

\section*{Acknowledgements}
We would like to thank IMPRS-PTFS for organizing the retreat which led to discussions shaping this paper and the support for TC. 
\bibliography{references.bib}

\begin{thebibliography}{57}%
\makeatletter
\providecommand \@ifxundefined [1]{%
 \@ifx{#1\undefined}
}%
\providecommand \@ifnum [1]{%
 \ifnum #1\expandafter \@firstoftwo
 \else \expandafter \@secondoftwo
 \fi
}%
\providecommand \@ifx [1]{%
 \ifx #1\expandafter \@firstoftwo
 \else \expandafter \@secondoftwo
 \fi
}%
\providecommand \natexlab [1]{#1}%
\providecommand \enquote  [1]{``#1''}%
\providecommand \bibnamefont  [1]{#1}%
\providecommand \bibfnamefont [1]{#1}%
\providecommand \citenamefont [1]{#1}%
\providecommand \href@noop [0]{\@secondoftwo}%
\providecommand \href [0]{\begingroup \@sanitize@url \@href}%
\providecommand \@href[1]{\@@startlink{#1}\@@href}%
\providecommand \@@href[1]{\endgroup#1\@@endlink}%
\providecommand \@sanitize@url [0]{\catcode `\\12\catcode `\$12\catcode
  `\&12\catcode `\#12\catcode `\^12\catcode `\_12\catcode `\%12\relax}%
\providecommand \@@startlink[1]{}%
\providecommand \@@endlink[0]{}%
\providecommand \url  [0]{\begingroup\@sanitize@url \@url }%
\providecommand \@url [1]{\endgroup\@href {#1}{\urlprefix }}%
\providecommand \urlprefix  [0]{URL }%
\providecommand \Eprint [0]{\href }%
\providecommand \doibase [0]{http://dx.doi.org/}%
\providecommand \selectlanguage [0]{\@gobble}%
\providecommand \bibinfo  [0]{\@secondoftwo}%
\providecommand \bibfield  [0]{\@secondoftwo}%
\providecommand \translation [1]{[#1]}%
\providecommand \BibitemOpen [0]{}%
\providecommand \bibitemStop [0]{}%
\providecommand \bibitemNoStop [0]{.\EOS\space}%
\providecommand \EOS [0]{\spacefactor3000\relax}%
\providecommand \BibitemShut  [1]{\csname bibitem#1\endcsname}%
\let\auto@bib@innerbib\@empty
\bibitem [{\citenamefont {Pauli}(1940)}]{Pauli:1940zz}%
  \BibitemOpen
  \bibfield  {author} {\bibinfo {author} {\bibfnamefont {W.}~\bibnamefont
  {Pauli}},\ }\href {\doibase 10.1103/PhysRev.58.716} {\bibfield  {journal}
  {\bibinfo  {journal} {Phys. Rev.}\ }\textbf {\bibinfo {volume} {58}},\
  \bibinfo {pages} {716} (\bibinfo {year} {1940})}\BibitemShut {NoStop}%
\bibitem [{\citenamefont {Luders}(1954)}]{Luders:1954zz}%
  \BibitemOpen
  \bibfield  {author} {\bibinfo {author} {\bibfnamefont {G.}~\bibnamefont
  {Luders}},\ }\href@noop {} {\bibfield  {journal} {\bibinfo  {journal} {Kong.
  Dan. Vid. Sel. Mat. Fys. Med.}\ }\textbf {\bibinfo {volume} {28N5}},\
  \bibinfo {pages} {1} (\bibinfo {year} {1954})}\BibitemShut {NoStop}%
\bibitem [{\citenamefont {Jost}(1957)}]{Jost:1957zz}%
  \BibitemOpen
  \bibfield  {author} {\bibinfo {author} {\bibfnamefont {R.}~\bibnamefont
  {Jost}},\ }\href@noop {} {\bibfield  {journal} {\bibinfo  {journal} {Helv.
  Phys. Acta}\ }\textbf {\bibinfo {volume} {30}},\ \bibinfo {pages} {409}
  (\bibinfo {year} {1957})}\BibitemShut {NoStop}%
\bibitem [{\citenamefont {Sakharov}(1967)}]{Sakharov:1967dj}%
  \BibitemOpen
  \bibfield  {author} {\bibinfo {author} {\bibfnamefont {A.~D.}\ \bibnamefont
  {Sakharov}},\ }\href {\doibase 10.1070/PU1991v034n05ABEH002497} {\bibfield
  {journal} {\bibinfo  {journal} {Pisma Zh. Eksp. Teor. Fiz.}\ }\textbf
  {\bibinfo {volume} {5}},\ \bibinfo {pages} {32} (\bibinfo {year}
  {1967})}\BibitemShut {NoStop}%
\bibitem [{\citenamefont {Bertolami}\ \emph {et~al.}(1997)\citenamefont
  {Bertolami}, \citenamefont {Colladay}, \citenamefont {Kostelecky},\ and\
  \citenamefont {Potting}}]{Bertolami:1996cq}%
  \BibitemOpen
  \bibfield  {author} {\bibinfo {author} {\bibfnamefont {O.}~\bibnamefont
  {Bertolami}}, \bibinfo {author} {\bibfnamefont {D.}~\bibnamefont {Colladay}},
  \bibinfo {author} {\bibfnamefont {V.~A.}\ \bibnamefont {Kostelecky}}, \ and\
  \bibinfo {author} {\bibfnamefont {R.}~\bibnamefont {Potting}},\ }\href
  {\doibase 10.1016/S0370-2693(97)00062-2} {\bibfield  {journal} {\bibinfo
  {journal} {Phys. Lett. B}\ }\textbf {\bibinfo {volume} {395}},\ \bibinfo
  {pages} {178} (\bibinfo {year} {1997})},\ \Eprint
  {http://arxiv.org/abs/hep-ph/9612437} {arXiv:hep-ph/9612437} \BibitemShut
  {NoStop}%
\bibitem [{\citenamefont {Tsujikawa}(2013)}]{Tsujikawa:2013fta}%
  \BibitemOpen
  \bibfield  {author} {\bibinfo {author} {\bibfnamefont {S.}~\bibnamefont
  {Tsujikawa}},\ }\href {\doibase 10.1088/0264-9381/30/21/214003} {\bibfield
  {journal} {\bibinfo  {journal} {Class. Quant. Grav.}\ }\textbf {\bibinfo
  {volume} {30}},\ \bibinfo {pages} {214003} (\bibinfo {year} {2013})},\
  \Eprint {http://arxiv.org/abs/1304.1961} {arXiv:1304.1961 [gr-qc]}
  \BibitemShut {NoStop}%
\bibitem [{\citenamefont {Colladay}\ and\ \citenamefont
  {Kostelecky}(1997)}]{Colladay:1996iz}%
  \BibitemOpen
  \bibfield  {author} {\bibinfo {author} {\bibfnamefont {D.}~\bibnamefont
  {Colladay}}\ and\ \bibinfo {author} {\bibfnamefont {V.~A.}\ \bibnamefont
  {Kostelecky}},\ }\href {\doibase 10.1103/PhysRevD.55.6760} {\bibfield
  {journal} {\bibinfo  {journal} {Phys. Rev. D}\ }\textbf {\bibinfo {volume}
  {55}},\ \bibinfo {pages} {6760} (\bibinfo {year} {1997})},\ \Eprint
  {http://arxiv.org/abs/hep-ph/9703464} {arXiv:hep-ph/9703464} \BibitemShut
  {NoStop}%
\bibitem [{\citenamefont {Colladay}\ and\ \citenamefont
  {Kostelecky}(1998)}]{Colladay:1998fq}%
  \BibitemOpen
  \bibfield  {author} {\bibinfo {author} {\bibfnamefont {D.}~\bibnamefont
  {Colladay}}\ and\ \bibinfo {author} {\bibfnamefont {V.~A.}\ \bibnamefont
  {Kostelecky}},\ }\href {\doibase 10.1103/PhysRevD.58.116002} {\bibfield
  {journal} {\bibinfo  {journal} {Phys. Rev. D}\ }\textbf {\bibinfo {volume}
  {58}},\ \bibinfo {pages} {116002} (\bibinfo {year} {1998})},\ \Eprint
  {http://arxiv.org/abs/hep-ph/9809521} {arXiv:hep-ph/9809521} \BibitemShut
  {NoStop}%
\bibitem [{\citenamefont {Kostelecky}\ and\ \citenamefont
  {Lehnert}(2001)}]{Kostelecky:2000mm}%
  \BibitemOpen
  \bibfield  {author} {\bibinfo {author} {\bibfnamefont {V.~A.}\ \bibnamefont
  {Kostelecky}}\ and\ \bibinfo {author} {\bibfnamefont {R.}~\bibnamefont
  {Lehnert}},\ }\href {\doibase 10.1103/PhysRevD.63.065008} {\bibfield
  {journal} {\bibinfo  {journal} {Phys. Rev. D}\ }\textbf {\bibinfo {volume}
  {63}},\ \bibinfo {pages} {065008} (\bibinfo {year} {2001})},\ \Eprint
  {http://arxiv.org/abs/hep-th/0012060} {arXiv:hep-th/0012060} \BibitemShut
  {NoStop}%
\bibitem [{\citenamefont {Coleman}\ and\ \citenamefont
  {Glashow}(1997)}]{Coleman:1997xq}%
  \BibitemOpen
  \bibfield  {author} {\bibinfo {author} {\bibfnamefont {S.~R.}\ \bibnamefont
  {Coleman}}\ and\ \bibinfo {author} {\bibfnamefont {S.~L.}\ \bibnamefont
  {Glashow}},\ }\href {\doibase 10.1016/S0370-2693(97)00638-2} {\bibfield
  {journal} {\bibinfo  {journal} {Phys. Lett. B}\ }\textbf {\bibinfo {volume}
  {405}},\ \bibinfo {pages} {249} (\bibinfo {year} {1997})},\ \Eprint
  {http://arxiv.org/abs/hep-ph/9703240} {arXiv:hep-ph/9703240} \BibitemShut
  {NoStop}%
\bibitem [{\citenamefont {Charlton}\ \emph {et~al.}(2020)\citenamefont
  {Charlton}, \citenamefont {Eriksson},\ and\ \citenamefont
  {Shore}}]{Charlton:2020xxc}%
  \BibitemOpen
  \bibfield  {author} {\bibinfo {author} {\bibfnamefont {M.}~\bibnamefont
  {Charlton}}, \bibinfo {author} {\bibfnamefont {S.}~\bibnamefont {Eriksson}},
  \ and\ \bibinfo {author} {\bibfnamefont {G.~M.}\ \bibnamefont {Shore}},\
  }\href@noop {} {\  (\bibinfo {year} {2020})},\ \Eprint
  {http://arxiv.org/abs/2002.09348} {arXiv:2002.09348 [hep-ph]} \BibitemShut
  {NoStop}%
\bibitem [{\citenamefont {Chaichian}\ \emph {et~al.}(2011)\citenamefont
  {Chaichian}, \citenamefont {Dolgov}, \citenamefont {Novikov},\ and\
  \citenamefont {Tureanu}}]{Chaichian:2011fc}%
  \BibitemOpen
  \bibfield  {author} {\bibinfo {author} {\bibfnamefont {M.}~\bibnamefont
  {Chaichian}}, \bibinfo {author} {\bibfnamefont {A.~D.}\ \bibnamefont
  {Dolgov}}, \bibinfo {author} {\bibfnamefont {V.~A.}\ \bibnamefont {Novikov}},
  \ and\ \bibinfo {author} {\bibfnamefont {A.}~\bibnamefont {Tureanu}},\ }\href
  {\doibase 10.1016/j.physletb.2011.03.026} {\bibfield  {journal} {\bibinfo
  {journal} {Phys. Lett. B}\ }\textbf {\bibinfo {volume} {699}},\ \bibinfo
  {pages} {177} (\bibinfo {year} {2011})},\ \Eprint
  {http://arxiv.org/abs/1103.0168} {arXiv:1103.0168 [hep-th]} \BibitemShut
  {NoStop}%
\bibitem [{\citenamefont {Dolgov}\ and\ \citenamefont
  {Novikov}(2012)}]{Dolgov:2012cm}%
  \BibitemOpen
  \bibfield  {author} {\bibinfo {author} {\bibfnamefont {A.~D.}\ \bibnamefont
  {Dolgov}}\ and\ \bibinfo {author} {\bibfnamefont {V.~A.}\ \bibnamefont
  {Novikov}},\ }\href {\doibase 10.1134/S0021364012110033} {\bibfield
  {journal} {\bibinfo  {journal} {JETP Lett.}\ }\textbf {\bibinfo {volume}
  {95}},\ \bibinfo {pages} {594} (\bibinfo {year} {2012})},\ \Eprint
  {http://arxiv.org/abs/1204.5612} {arXiv:1204.5612 [hep-ph]} \BibitemShut
  {NoStop}%
\bibitem [{\citenamefont {Chaichian}\ \emph
  {et~al.}(2013{\natexlab{a}})\citenamefont {Chaichian}, \citenamefont
  {Fujikawa},\ and\ \citenamefont {Tureanu}}]{Chaichian:2012hy}%
  \BibitemOpen
  \bibfield  {author} {\bibinfo {author} {\bibfnamefont {M.}~\bibnamefont
  {Chaichian}}, \bibinfo {author} {\bibfnamefont {K.}~\bibnamefont {Fujikawa}},
  \ and\ \bibinfo {author} {\bibfnamefont {A.}~\bibnamefont {Tureanu}},\ }\href
  {\doibase 10.1016/j.physletb.2012.12.017} {\bibfield  {journal} {\bibinfo
  {journal} {Phys. Lett. B}\ }\textbf {\bibinfo {volume} {718}},\ \bibinfo
  {pages} {1500} (\bibinfo {year} {2013}{\natexlab{a}})},\ \Eprint
  {http://arxiv.org/abs/1210.0208} {arXiv:1210.0208 [hep-th]} \BibitemShut
  {NoStop}%
\bibitem [{\citenamefont {Fujikawa}\ and\ \citenamefont
  {Tureanu}(2017)}]{Fujikawa:2016her}%
  \BibitemOpen
  \bibfield  {author} {\bibinfo {author} {\bibfnamefont {K.}~\bibnamefont
  {Fujikawa}}\ and\ \bibinfo {author} {\bibfnamefont {A.}~\bibnamefont
  {Tureanu}},\ }\href {\doibase 10.1142/S0217751X17410147} {\bibfield
  {journal} {\bibinfo  {journal} {Int. J. Mod. Phys. A}\ }\textbf {\bibinfo
  {volume} {32}},\ \bibinfo {pages} {1741014} (\bibinfo {year} {2017})},\
  \Eprint {http://arxiv.org/abs/1607.01409} {arXiv:1607.01409 [hep-ph]}
  \BibitemShut {NoStop}%
\bibitem [{BAR(1984)}]{BARMIN1984293}%
  \BibitemOpen
  \href@noop {} {\bibfield  {journal} {\bibinfo  {journal} {Nuclear Physics B}\
  }\textbf {\bibinfo {volume} {247}},\ \bibinfo {pages} {293} (\bibinfo {year}
  {1984})}\BibitemShut {NoStop}%
\bibitem [{\citenamefont {Workman}\ and\ \citenamefont
  {Others}(2022)}]{Workman:2022ynf}%
  \BibitemOpen
  \bibfield  {author} {\bibinfo {author} {\bibfnamefont {R.~L.}\ \bibnamefont
  {Workman}}\ and\ \bibinfo {author} {\bibnamefont {Others}} (\bibinfo
  {collaboration} {Particle Data Group}),\ }\href {\doibase
  10.1093/ptep/ptac097} {\bibfield  {journal} {\bibinfo  {journal} {PTEP}\
  }\textbf {\bibinfo {volume} {2022}},\ \bibinfo {pages} {083C01} (\bibinfo
  {year} {2022})}\BibitemShut {NoStop}%
\bibitem [{\citenamefont {Nierste}(2009)}]{Nierste:2009wg}%
  \BibitemOpen
  \bibfield  {author} {\bibinfo {author} {\bibfnamefont {U.}~\bibnamefont
  {Nierste}},\ }in\ \href@noop {} {\emph {\bibinfo {booktitle} {{Helmholz
  International Summer School on Heavy Quark Physics}}}}\ (\bibinfo {year}
  {2009})\ pp.\ \bibinfo {pages} {1--38},\ \Eprint
  {http://arxiv.org/abs/0904.1869} {arXiv:0904.1869 [hep-ph]} \BibitemShut
  {NoStop}%
\bibitem [{\citenamefont {De~Gouvea}(2002)}]{DeGouvea:2002xp}%
  \BibitemOpen
  \bibfield  {author} {\bibinfo {author} {\bibfnamefont {A.}~\bibnamefont
  {De~Gouvea}},\ }\href {\doibase 10.1103/PhysRevD.66.076005} {\bibfield
  {journal} {\bibinfo  {journal} {Phys. Rev. D}\ }\textbf {\bibinfo {volume}
  {66}},\ \bibinfo {pages} {076005} (\bibinfo {year} {2002})},\ \Eprint
  {http://arxiv.org/abs/hep-ph/0204077} {arXiv:hep-ph/0204077} \BibitemShut
  {NoStop}%
\bibitem [{\citenamefont {Bahcall}\ \emph {et~al.}(2002)\citenamefont
  {Bahcall}, \citenamefont {Barger},\ and\ \citenamefont
  {Marfatia}}]{Bahcall:2002ia}%
  \BibitemOpen
  \bibfield  {author} {\bibinfo {author} {\bibfnamefont {J.~N.}\ \bibnamefont
  {Bahcall}}, \bibinfo {author} {\bibfnamefont {V.}~\bibnamefont {Barger}}, \
  and\ \bibinfo {author} {\bibfnamefont {D.}~\bibnamefont {Marfatia}},\ }\href
  {\doibase 10.1016/S0370-2693(02)01714-8} {\bibfield  {journal} {\bibinfo
  {journal} {Phys. Lett. B}\ }\textbf {\bibinfo {volume} {534}},\ \bibinfo
  {pages} {120} (\bibinfo {year} {2002})},\ \Eprint
  {http://arxiv.org/abs/hep-ph/0201211} {arXiv:hep-ph/0201211} \BibitemShut
  {NoStop}%
\bibitem [{\citenamefont {Kostelecky}\ and\ \citenamefont
  {Mewes}(2004{\natexlab{a}})}]{Kostelecky:2003xn}%
  \BibitemOpen
  \bibfield  {author} {\bibinfo {author} {\bibfnamefont {V.~A.}\ \bibnamefont
  {Kostelecky}}\ and\ \bibinfo {author} {\bibfnamefont {M.}~\bibnamefont
  {Mewes}},\ }\href {\doibase 10.1103/PhysRevD.70.031902} {\bibfield  {journal}
  {\bibinfo  {journal} {Phys. Rev. D}\ }\textbf {\bibinfo {volume} {70}},\
  \bibinfo {pages} {031902} (\bibinfo {year} {2004}{\natexlab{a}})},\ \Eprint
  {http://arxiv.org/abs/hep-ph/0308300} {arXiv:hep-ph/0308300} \BibitemShut
  {NoStop}%
\bibitem [{\citenamefont {Kostelecky}\ and\ \citenamefont
  {Mewes}(2004{\natexlab{b}})}]{Kostelecky:2003cr}%
  \BibitemOpen
  \bibfield  {author} {\bibinfo {author} {\bibfnamefont {V.~A.}\ \bibnamefont
  {Kostelecky}}\ and\ \bibinfo {author} {\bibfnamefont {M.}~\bibnamefont
  {Mewes}},\ }\href {\doibase 10.1103/PhysRevD.69.016005} {\bibfield  {journal}
  {\bibinfo  {journal} {Phys. Rev. D}\ }\textbf {\bibinfo {volume} {69}},\
  \bibinfo {pages} {016005} (\bibinfo {year} {2004}{\natexlab{b}})},\ \Eprint
  {http://arxiv.org/abs/hep-ph/0309025} {arXiv:hep-ph/0309025} \BibitemShut
  {NoStop}%
\bibitem [{\citenamefont {Datta}\ \emph {et~al.}(2004)\citenamefont {Datta},
  \citenamefont {Gandhi}, \citenamefont {Mehta},\ and\ \citenamefont
  {Sankar}}]{Datta:2003dg}%
  \BibitemOpen
  \bibfield  {author} {\bibinfo {author} {\bibfnamefont {A.}~\bibnamefont
  {Datta}}, \bibinfo {author} {\bibfnamefont {R.}~\bibnamefont {Gandhi}},
  \bibinfo {author} {\bibfnamefont {P.}~\bibnamefont {Mehta}}, \ and\ \bibinfo
  {author} {\bibfnamefont {S.~U.}\ \bibnamefont {Sankar}},\ }\href {\doibase
  10.1016/j.physletb.2004.07.035} {\bibfield  {journal} {\bibinfo  {journal}
  {Phys. Lett. B}\ }\textbf {\bibinfo {volume} {597}},\ \bibinfo {pages} {356}
  (\bibinfo {year} {2004})},\ \Eprint {http://arxiv.org/abs/hep-ph/0312027}
  {arXiv:hep-ph/0312027} \BibitemShut {NoStop}%
\bibitem [{\citenamefont {Minakata}\ and\ \citenamefont
  {Uchinami}(2005)}]{Minakata:2005jy}%
  \BibitemOpen
  \bibfield  {author} {\bibinfo {author} {\bibfnamefont {H.}~\bibnamefont
  {Minakata}}\ and\ \bibinfo {author} {\bibfnamefont {S.}~\bibnamefont
  {Uchinami}},\ }\href {\doibase 10.1103/PhysRevD.72.105007} {\bibfield
  {journal} {\bibinfo  {journal} {Phys. Rev. D}\ }\textbf {\bibinfo {volume}
  {72}},\ \bibinfo {pages} {105007} (\bibinfo {year} {2005})},\ \Eprint
  {http://arxiv.org/abs/hep-ph/0505133} {arXiv:hep-ph/0505133} \BibitemShut
  {NoStop}%
\bibitem [{\citenamefont {Kostelecky}\ and\ \citenamefont
  {Mewes}(2012)}]{Kostelecky:2011gq}%
  \BibitemOpen
  \bibfield  {author} {\bibinfo {author} {\bibfnamefont {A.}~\bibnamefont
  {Kostelecky}}\ and\ \bibinfo {author} {\bibfnamefont {M.}~\bibnamefont
  {Mewes}},\ }\href {\doibase 10.1103/PhysRevD.85.096005} {\bibfield  {journal}
  {\bibinfo  {journal} {Phys. Rev. D}\ }\textbf {\bibinfo {volume} {85}},\
  \bibinfo {pages} {096005} (\bibinfo {year} {2012})},\ \Eprint
  {http://arxiv.org/abs/1112.6395} {arXiv:1112.6395 [hep-ph]} \BibitemShut
  {NoStop}%
\bibitem [{\citenamefont {Ohlsson}\ and\ \citenamefont
  {Zhou}(2015)}]{Ohlsson:2014cha}%
  \BibitemOpen
  \bibfield  {author} {\bibinfo {author} {\bibfnamefont {T.}~\bibnamefont
  {Ohlsson}}\ and\ \bibinfo {author} {\bibfnamefont {S.}~\bibnamefont {Zhou}},\
  }\href {\doibase 10.1016/j.nuclphysb.2015.02.015} {\bibfield  {journal}
  {\bibinfo  {journal} {Nucl. Phys. B}\ }\textbf {\bibinfo {volume} {893}},\
  \bibinfo {pages} {482} (\bibinfo {year} {2015})},\ \Eprint
  {http://arxiv.org/abs/1408.4722} {arXiv:1408.4722 [hep-ph]} \BibitemShut
  {NoStop}%
\bibitem [{\citenamefont {Abe}\ \emph {et~al.}(2015)\citenamefont {Abe} \emph
  {et~al.}}]{Super-Kamiokande:2014exs}%
  \BibitemOpen
  \bibfield  {author} {\bibinfo {author} {\bibfnamefont {K.}~\bibnamefont
  {Abe}} \emph {et~al.} (\bibinfo {collaboration} {Super-Kamiokande}),\ }\href
  {\doibase 10.1103/PhysRevD.91.052003} {\bibfield  {journal} {\bibinfo
  {journal} {Phys. Rev. D}\ }\textbf {\bibinfo {volume} {91}},\ \bibinfo
  {pages} {052003} (\bibinfo {year} {2015})},\ \Eprint
  {http://arxiv.org/abs/1410.4267} {arXiv:1410.4267 [hep-ex]} \BibitemShut
  {NoStop}%
\bibitem [{\citenamefont {Diaz}\ and\ \citenamefont
  {Schwetz}(2016)}]{Diaz:2016fqd}%
  \BibitemOpen
  \bibfield  {author} {\bibinfo {author} {\bibfnamefont {J.~S.}\ \bibnamefont
  {Diaz}}\ and\ \bibinfo {author} {\bibfnamefont {T.}~\bibnamefont {Schwetz}},\
  }\href {\doibase 10.1103/PhysRevD.93.093004} {\bibfield  {journal} {\bibinfo
  {journal} {Phys. Rev. D}\ }\textbf {\bibinfo {volume} {93}},\ \bibinfo
  {pages} {093004} (\bibinfo {year} {2016})},\ \Eprint
  {http://arxiv.org/abs/1603.04468} {arXiv:1603.04468 [hep-ph]} \BibitemShut
  {NoStop}%
\bibitem [{\citenamefont {de~Gouv\^ea}\ and\ \citenamefont
  {Kelly}(2017)}]{deGouvea:2017yvn}%
  \BibitemOpen
  \bibfield  {author} {\bibinfo {author} {\bibfnamefont {A.}~\bibnamefont
  {de~Gouv\^ea}}\ and\ \bibinfo {author} {\bibfnamefont {K.~J.}\ \bibnamefont
  {Kelly}},\ }\href {\doibase 10.1103/PhysRevD.96.095018} {\bibfield  {journal}
  {\bibinfo  {journal} {Phys. Rev. D}\ }\textbf {\bibinfo {volume} {96}},\
  \bibinfo {pages} {095018} (\bibinfo {year} {2017})},\ \Eprint
  {http://arxiv.org/abs/1709.06090} {arXiv:1709.06090 [hep-ph]} \BibitemShut
  {NoStop}%
\bibitem [{\citenamefont {Barenboim}\ \emph {et~al.}(2018)\citenamefont
  {Barenboim}, \citenamefont {Ternes},\ and\ \citenamefont
  {T\'ortola}}]{Barenboim:2017ewj}%
  \BibitemOpen
  \bibfield  {author} {\bibinfo {author} {\bibfnamefont {G.}~\bibnamefont
  {Barenboim}}, \bibinfo {author} {\bibfnamefont {C.~A.}\ \bibnamefont
  {Ternes}}, \ and\ \bibinfo {author} {\bibfnamefont {M.}~\bibnamefont
  {T\'ortola}},\ }\href {\doibase 10.1016/j.physletb.2018.03.060} {\bibfield
  {journal} {\bibinfo  {journal} {Phys. Lett. B}\ }\textbf {\bibinfo {volume}
  {780}},\ \bibinfo {pages} {631} (\bibinfo {year} {2018})},\ \Eprint
  {http://arxiv.org/abs/1712.01714} {arXiv:1712.01714 [hep-ph]} \BibitemShut
  {NoStop}%
\bibitem [{\citenamefont {Liberati}(2013)}]{Liberati:2013xla}%
  \BibitemOpen
  \bibfield  {author} {\bibinfo {author} {\bibfnamefont {S.}~\bibnamefont
  {Liberati}},\ }\href {\doibase 10.1088/0264-9381/30/13/133001} {\bibfield
  {journal} {\bibinfo  {journal} {Class. Quant. Grav.}\ }\textbf {\bibinfo
  {volume} {30}},\ \bibinfo {pages} {133001} (\bibinfo {year} {2013})},\
  \Eprint {http://arxiv.org/abs/1304.5795} {arXiv:1304.5795 [gr-qc]}
  \BibitemShut {NoStop}%
\bibitem [{\citenamefont {Antusch}\ and\ \citenamefont
  {Fernandez-Martinez}(2008)}]{Antusch:2008zj}%
  \BibitemOpen
  \bibfield  {author} {\bibinfo {author} {\bibfnamefont {S.}~\bibnamefont
  {Antusch}}\ and\ \bibinfo {author} {\bibfnamefont {E.}~\bibnamefont
  {Fernandez-Martinez}},\ }\href {\doibase 10.1016/j.physletb.2008.06.005}
  {\bibfield  {journal} {\bibinfo  {journal} {Phys. Lett. B}\ }\textbf
  {\bibinfo {volume} {665}},\ \bibinfo {pages} {190} (\bibinfo {year}
  {2008})},\ \Eprint {http://arxiv.org/abs/0804.2820} {arXiv:0804.2820
  [hep-ph]} \BibitemShut {NoStop}%
\bibitem [{\citenamefont {Amole}\ \emph {et~al.}(2014)\citenamefont {Amole}
  \emph {et~al.}}]{ALPHA:2014nwo}%
  \BibitemOpen
  \bibfield  {author} {\bibinfo {author} {\bibfnamefont {C.}~\bibnamefont
  {Amole}} \emph {et~al.} (\bibinfo {collaboration} {ALPHA}),\ }\href {\doibase
  10.1038/ncomms4955} {\bibfield  {journal} {\bibinfo  {journal} {Nature
  Commun.}\ }\textbf {\bibinfo {volume} {5}},\ \bibinfo {pages} {3955}
  (\bibinfo {year} {2014})}\BibitemShut {NoStop}%
\bibitem [{\citenamefont {Ahmadi}\ \emph {et~al.}(2016)\citenamefont {Ahmadi}
  \emph {et~al.}}]{ALPHA:2016klx}%
  \BibitemOpen
  \bibfield  {author} {\bibinfo {author} {\bibfnamefont {M.}~\bibnamefont
  {Ahmadi}} \emph {et~al.} (\bibinfo {collaboration} {ALPHA}),\ }\href
  {\doibase 10.1038/nature16491} {\bibfield  {journal} {\bibinfo  {journal}
  {Nature}\ }\textbf {\bibinfo {volume} {529}},\ \bibinfo {pages} {373}
  (\bibinfo {year} {2016})}\BibitemShut {NoStop}%
\bibitem [{\citenamefont {Bertsche}(2018)}]{Bertsche:2018avg}%
  \BibitemOpen
  \bibfield  {author} {\bibinfo {author} {\bibfnamefont {W.~A.}\ \bibnamefont
  {Bertsche}},\ }\href {\doibase 10.1098/rsta.2017.0265} {\bibfield  {journal}
  {\bibinfo  {journal} {Phil. Trans. Roy. Soc. Lond. A}\ }\textbf {\bibinfo
  {volume} {376}},\ \bibinfo {pages} {20170265} (\bibinfo {year}
  {2018})}\BibitemShut {NoStop}%
\bibitem [{\citenamefont {Ahmadi}\ \emph {et~al.}(2017)\citenamefont {Ahmadi}
  \emph {et~al.}}]{ALPHA:2017fsh}%
  \BibitemOpen
  \bibfield  {author} {\bibinfo {author} {\bibfnamefont {M.}~\bibnamefont
  {Ahmadi}} \emph {et~al.} (\bibinfo {collaboration} {ALPHA}),\ }\href
  {\doibase 10.1038/nature23446} {\bibfield  {journal} {\bibinfo  {journal}
  {Nature}\ }\textbf {\bibinfo {volume} {548}},\ \bibinfo {pages} {66}
  (\bibinfo {year} {2017})}\BibitemShut {NoStop}%
\bibitem [{\citenamefont {Brown}\ and\ \citenamefont
  {Gabrielse}(1986)}]{Brown:1985rh}%
  \BibitemOpen
  \bibfield  {author} {\bibinfo {author} {\bibfnamefont {L.~S.}\ \bibnamefont
  {Brown}}\ and\ \bibinfo {author} {\bibfnamefont {G.}~\bibnamefont
  {Gabrielse}},\ }\href {\doibase 10.1103/RevModPhys.58.233} {\bibfield
  {journal} {\bibinfo  {journal} {Rev. Mod. Phys.}\ }\textbf {\bibinfo {volume}
  {58}},\ \bibinfo {pages} {233} (\bibinfo {year} {1986})}\BibitemShut
  {NoStop}%
\bibitem [{\citenamefont {Borchert}\ \emph {et~al.}(2022)\citenamefont
  {Borchert} \emph {et~al.}}]{BASE:2022yvh}%
  \BibitemOpen
  \bibfield  {author} {\bibinfo {author} {\bibfnamefont {M.~J.}\ \bibnamefont
  {Borchert}} \emph {et~al.} (\bibinfo {collaboration} {BASE}),\ }\href
  {\doibase 10.1038/s41586-021-04203-w} {\bibfield  {journal} {\bibinfo
  {journal} {Nature}\ }\textbf {\bibinfo {volume} {601}},\ \bibinfo {pages}
  {53} (\bibinfo {year} {2022})}\BibitemShut {NoStop}%
\bibitem [{\citenamefont {Hughes}\ and\ \citenamefont
  {Holzscheiter}(1991)}]{Hughes:1990ay}%
  \BibitemOpen
  \bibfield  {author} {\bibinfo {author} {\bibfnamefont {R.~J.}\ \bibnamefont
  {Hughes}}\ and\ \bibinfo {author} {\bibfnamefont {M.~H.}\ \bibnamefont
  {Holzscheiter}},\ }\href {\doibase 10.1103/PhysRevLett.66.854} {\bibfield
  {journal} {\bibinfo  {journal} {Phys. Rev. Lett.}\ }\textbf {\bibinfo
  {volume} {66}},\ \bibinfo {pages} {854} (\bibinfo {year} {1991})}\BibitemShut
  {NoStop}%
\bibitem [{\citenamefont {Caldwell}\ and\ \citenamefont
  {Dvali}(2021)}]{Caldwell:2019icl}%
  \BibitemOpen
  \bibfield  {author} {\bibinfo {author} {\bibfnamefont {A.}~\bibnamefont
  {Caldwell}}\ and\ \bibinfo {author} {\bibfnamefont {G.}~\bibnamefont
  {Dvali}},\ }\href {\doibase 10.1002/prop.202000092} {\bibfield  {journal}
  {\bibinfo  {journal} {Fortsch. Phys.}\ }\textbf {\bibinfo {volume} {69}},\
  \bibinfo {pages} {2000092} (\bibinfo {year} {2021})},\ \Eprint
  {http://arxiv.org/abs/1903.09096} {arXiv:1903.09096 [hep-ph]} \BibitemShut
  {NoStop}%
\bibitem [{\citenamefont {Ji}(1995)}]{Ji:1994av}%
  \BibitemOpen
  \bibfield  {author} {\bibinfo {author} {\bibfnamefont {X.-D.}\ \bibnamefont
  {Ji}},\ }\href {\doibase 10.1103/PhysRevLett.74.1071} {\bibfield  {journal}
  {\bibinfo  {journal} {Phys. Rev. Lett.}\ }\textbf {\bibinfo {volume} {74}},\
  \bibinfo {pages} {1071} (\bibinfo {year} {1995})},\ \Eprint
  {http://arxiv.org/abs/hep-ph/9410274} {arXiv:hep-ph/9410274} \BibitemShut
  {NoStop}%
\bibitem [{\citenamefont {Yang}\ \emph {et~al.}(2015)\citenamefont {Yang},
  \citenamefont {Chen}, \citenamefont {Draper}, \citenamefont {Gong},
  \citenamefont {Liu}, \citenamefont {Liu},\ and\ \citenamefont
  {Ma}}]{Yang:2014xsa}%
  \BibitemOpen
  \bibfield  {author} {\bibinfo {author} {\bibfnamefont {Y.-B.}\ \bibnamefont
  {Yang}}, \bibinfo {author} {\bibfnamefont {Y.}~\bibnamefont {Chen}}, \bibinfo
  {author} {\bibfnamefont {T.}~\bibnamefont {Draper}}, \bibinfo {author}
  {\bibfnamefont {M.}~\bibnamefont {Gong}}, \bibinfo {author} {\bibfnamefont
  {K.-F.}\ \bibnamefont {Liu}}, \bibinfo {author} {\bibfnamefont
  {Z.}~\bibnamefont {Liu}}, \ and\ \bibinfo {author} {\bibfnamefont {J.-P.}\
  \bibnamefont {Ma}},\ }\href {\doibase 10.1103/PhysRevD.91.074516} {\bibfield
  {journal} {\bibinfo  {journal} {Phys. Rev. D}\ }\textbf {\bibinfo {volume}
  {91}},\ \bibinfo {pages} {074516} (\bibinfo {year} {2015})},\ \Eprint
  {http://arxiv.org/abs/1405.4440} {arXiv:1405.4440 [hep-ph]} \BibitemShut
  {NoStop}%
\bibitem [{\citenamefont {Yang}\ \emph {et~al.}(2018)\citenamefont {Yang},
  \citenamefont {Liang}, \citenamefont {Bi}, \citenamefont {Chen},
  \citenamefont {Draper}, \citenamefont {Liu},\ and\ \citenamefont
  {Liu}}]{Yang:2018nqn}%
  \BibitemOpen
  \bibfield  {author} {\bibinfo {author} {\bibfnamefont {Y.-B.}\ \bibnamefont
  {Yang}}, \bibinfo {author} {\bibfnamefont {J.}~\bibnamefont {Liang}},
  \bibinfo {author} {\bibfnamefont {Y.-J.}\ \bibnamefont {Bi}}, \bibinfo
  {author} {\bibfnamefont {Y.}~\bibnamefont {Chen}}, \bibinfo {author}
  {\bibfnamefont {T.}~\bibnamefont {Draper}}, \bibinfo {author} {\bibfnamefont
  {K.-F.}\ \bibnamefont {Liu}}, \ and\ \bibinfo {author} {\bibfnamefont
  {Z.}~\bibnamefont {Liu}},\ }\href {\doibase 10.1103/PhysRevLett.121.212001}
  {\bibfield  {journal} {\bibinfo  {journal} {Phys. Rev. Lett.}\ }\textbf
  {\bibinfo {volume} {121}},\ \bibinfo {pages} {212001} (\bibinfo {year}
  {2018})},\ \Eprint {http://arxiv.org/abs/1808.08677} {arXiv:1808.08677
  [hep-lat]} \BibitemShut {NoStop}%
\bibitem [{\citenamefont {Baikov}\ \emph {et~al.}(2014)\citenamefont {Baikov},
  \citenamefont {Chetyrkin},\ and\ \citenamefont {K\"uhn}}]{Baikov:2014qja}%
  \BibitemOpen
  \bibfield  {author} {\bibinfo {author} {\bibfnamefont {P.~A.}\ \bibnamefont
  {Baikov}}, \bibinfo {author} {\bibfnamefont {K.~G.}\ \bibnamefont
  {Chetyrkin}}, \ and\ \bibinfo {author} {\bibfnamefont {J.~H.}\ \bibnamefont
  {K\"uhn}},\ }\href {\doibase 10.1007/JHEP10(2014)076} {\bibfield  {journal}
  {\bibinfo  {journal} {JHEP}\ }\textbf {\bibinfo {volume} {10}},\ \bibinfo
  {pages} {076} (\bibinfo {year} {2014})},\ \Eprint
  {http://arxiv.org/abs/1402.6611} {arXiv:1402.6611 [hep-ph]} \BibitemShut
  {NoStop}%
\bibitem [{\citenamefont {Aoki}\ \emph {et~al.}(2022)\citenamefont {Aoki} \emph
  {et~al.}}]{FlavourLatticeAveragingGroupFLAG:2021npn}%
  \BibitemOpen
  \bibfield  {author} {\bibinfo {author} {\bibfnamefont {Y.}~\bibnamefont
  {Aoki}} \emph {et~al.} (\bibinfo {collaboration} {Flavour Lattice Averaging
  Group (FLAG)}),\ }\href {\doibase 10.1140/epjc/s10052-022-10536-1} {\bibfield
   {journal} {\bibinfo  {journal} {Eur. Phys. J. C}\ }\textbf {\bibinfo
  {volume} {82}},\ \bibinfo {pages} {869} (\bibinfo {year} {2022})},\ \Eprint
  {http://arxiv.org/abs/2111.09849} {arXiv:2111.09849 [hep-lat]} \BibitemShut
  {NoStop}%
\bibitem [{\citenamefont {Gupta}\ \emph {et~al.}(2021)\citenamefont {Gupta},
  \citenamefont {Park}, \citenamefont {Hoferichter}, \citenamefont
  {Mereghetti}, \citenamefont {Yoon},\ and\ \citenamefont
  {Bhattacharya}}]{Gupta:2021ahb}%
  \BibitemOpen
  \bibfield  {author} {\bibinfo {author} {\bibfnamefont {R.}~\bibnamefont
  {Gupta}}, \bibinfo {author} {\bibfnamefont {S.}~\bibnamefont {Park}},
  \bibinfo {author} {\bibfnamefont {M.}~\bibnamefont {Hoferichter}}, \bibinfo
  {author} {\bibfnamefont {E.}~\bibnamefont {Mereghetti}}, \bibinfo {author}
  {\bibfnamefont {B.}~\bibnamefont {Yoon}}, \ and\ \bibinfo {author}
  {\bibfnamefont {T.}~\bibnamefont {Bhattacharya}},\ }\href {\doibase
  10.1103/PhysRevLett.127.242002} {\bibfield  {journal} {\bibinfo  {journal}
  {Phys. Rev. Lett.}\ }\textbf {\bibinfo {volume} {127}},\ \bibinfo {pages}
  {242002} (\bibinfo {year} {2021})},\ \Eprint
  {http://arxiv.org/abs/2105.12095} {arXiv:2105.12095 [hep-lat]} \BibitemShut
  {NoStop}%
\bibitem [{\citenamefont {Bell}\ and\ \citenamefont
  {Steinberger}(1966)}]{Bell:1966vvu}%
  \BibitemOpen
  \bibfield  {author} {\bibinfo {author} {\bibfnamefont {J.~S.}\ \bibnamefont
  {Bell}}\ and\ \bibinfo {author} {\bibfnamefont {J.}~\bibnamefont
  {Steinberger}},\ }in\ \href@noop {} {\emph {\bibinfo {booktitle} {{Oxford
  International Conference on Elementary Particles}}}}\ (\bibinfo {year}
  {1966})\ pp.\ \bibinfo {pages} {195--222}\BibitemShut {NoStop}%
\bibitem [{\citenamefont {Aguilar}\ \emph {et~al.}(2019)\citenamefont {Aguilar}
  \emph {et~al.}}]{Aguilar:2019teb}%
  \BibitemOpen
  \bibfield  {author} {\bibinfo {author} {\bibfnamefont {A.~C.}\ \bibnamefont
  {Aguilar}} \emph {et~al.},\ }\href {\doibase 10.1140/epja/i2019-12885-0}
  {\bibfield  {journal} {\bibinfo  {journal} {Eur. Phys. J. A}\ }\textbf
  {\bibinfo {volume} {55}},\ \bibinfo {pages} {190} (\bibinfo {year} {2019})},\
  \Eprint {http://arxiv.org/abs/1907.08218} {arXiv:1907.08218 [nucl-ex]}
  \BibitemShut {NoStop}%
\bibitem [{\citenamefont {T\'ortola}\ \emph {et~al.}(2020)\citenamefont
  {T\'ortola}, \citenamefont {Barenboim},\ and\ \citenamefont
  {Ternes}}]{Tortola:2020ncu}%
  \BibitemOpen
  \bibfield  {author} {\bibinfo {author} {\bibfnamefont {M.~A.}\ \bibnamefont
  {T\'ortola}}, \bibinfo {author} {\bibfnamefont {G.}~\bibnamefont
  {Barenboim}}, \ and\ \bibinfo {author} {\bibfnamefont {C.~A.}\ \bibnamefont
  {Ternes}},\ }\href {\doibase 10.1007/JHEP07(2020)155} {\bibfield  {journal}
  {\bibinfo  {journal} {JHEP}\ }\textbf {\bibinfo {volume} {07}},\ \bibinfo
  {pages} {155} (\bibinfo {year} {2020})},\ \Eprint
  {http://arxiv.org/abs/2005.05975} {arXiv:2005.05975 [hep-ph]} \BibitemShut
  {NoStop}%
\bibitem [{\citenamefont {Murayama}(2004)}]{Murayama:2003zw}%
  \BibitemOpen
  \bibfield  {author} {\bibinfo {author} {\bibfnamefont {H.}~\bibnamefont
  {Murayama}},\ }\href {\doibase 10.1016/j.physletb.2004.06.106} {\bibfield
  {journal} {\bibinfo  {journal} {Phys. Lett. B}\ }\textbf {\bibinfo {volume}
  {597}},\ \bibinfo {pages} {73} (\bibinfo {year} {2004})},\ \Eprint
  {http://arxiv.org/abs/hep-ph/0307127} {arXiv:hep-ph/0307127} \BibitemShut
  {NoStop}%
\bibitem [{\citenamefont {Tomboulis}(2015)}]{Tomboulis:2015gfa}%
  \BibitemOpen
  \bibfield  {author} {\bibinfo {author} {\bibfnamefont {E.~T.}\ \bibnamefont
  {Tomboulis}},\ }\href {\doibase 10.1103/PhysRevD.92.125037} {\bibfield
  {journal} {\bibinfo  {journal} {Phys. Rev. D}\ }\textbf {\bibinfo {volume}
  {92}},\ \bibinfo {pages} {125037} (\bibinfo {year} {2015})},\ \Eprint
  {http://arxiv.org/abs/1507.00981} {arXiv:1507.00981 [hep-th]} \BibitemShut
  {NoStop}%
\bibitem [{\citenamefont {Greenberg}(2002)}]{Greenberg:2002uu}%
  \BibitemOpen
  \bibfield  {author} {\bibinfo {author} {\bibfnamefont {O.~W.}\ \bibnamefont
  {Greenberg}},\ }\href {\doibase 10.1103/PhysRevLett.89.231602} {\bibfield
  {journal} {\bibinfo  {journal} {Phys. Rev. Lett.}\ }\textbf {\bibinfo
  {volume} {89}},\ \bibinfo {pages} {231602} (\bibinfo {year} {2002})},\
  \Eprint {http://arxiv.org/abs/hep-ph/0201258} {arXiv:hep-ph/0201258}
  \BibitemShut {NoStop}%
\bibitem [{\citenamefont {Chaichian}\ \emph
  {et~al.}(2013{\natexlab{b}})\citenamefont {Chaichian}, \citenamefont
  {Fujikawa},\ and\ \citenamefont {Tureanu}}]{Chaichian:2012ga}%
  \BibitemOpen
  \bibfield  {author} {\bibinfo {author} {\bibfnamefont {M.}~\bibnamefont
  {Chaichian}}, \bibinfo {author} {\bibfnamefont {K.}~\bibnamefont {Fujikawa}},
  \ and\ \bibinfo {author} {\bibfnamefont {A.}~\bibnamefont {Tureanu}},\ }\href
  {\doibase 10.1140/epjc/s10052-013-2349-2} {\bibfield  {journal} {\bibinfo
  {journal} {Eur. Phys. J. C}\ }\textbf {\bibinfo {volume} {73}},\ \bibinfo
  {pages} {2349} (\bibinfo {year} {2013}{\natexlab{b}})},\ \Eprint
  {http://arxiv.org/abs/1205.0152} {arXiv:1205.0152 [hep-th]} \BibitemShut
  {NoStop}%
\bibitem [{\citenamefont {Larin}(2020)}]{Larin:2020hih}%
  \BibitemOpen
  \bibfield  {author} {\bibinfo {author} {\bibfnamefont {S.~A.}\ \bibnamefont
  {Larin}},\ }\href {\doibase 10.1209/0295-5075/129/21003} {\bibfield
  {journal} {\bibinfo  {journal} {EPL}\ }\textbf {\bibinfo {volume} {129}},\
  \bibinfo {pages} {21003} (\bibinfo {year} {2020})},\ \Eprint
  {http://arxiv.org/abs/2006.00910} {arXiv:2006.00910 [hep-ph]} \BibitemShut
  {NoStop}%
\bibitem [{\citenamefont {Kostelecky}\ and\ \citenamefont
  {Samuel}(1989{\natexlab{a}})}]{Kostelecky:1988zi}%
  \BibitemOpen
  \bibfield  {author} {\bibinfo {author} {\bibfnamefont {V.~A.}\ \bibnamefont
  {Kostelecky}}\ and\ \bibinfo {author} {\bibfnamefont {S.}~\bibnamefont
  {Samuel}},\ }\href {\doibase 10.1103/PhysRevD.39.683} {\bibfield  {journal}
  {\bibinfo  {journal} {Phys. Rev. D}\ }\textbf {\bibinfo {volume} {39}},\
  \bibinfo {pages} {683} (\bibinfo {year} {1989}{\natexlab{a}})}\BibitemShut
  {NoStop}%
\bibitem [{\citenamefont {Kostelecky}\ and\ \citenamefont
  {Samuel}(1989{\natexlab{b}})}]{Kostelecky:1989jw}%
  \BibitemOpen
  \bibfield  {author} {\bibinfo {author} {\bibfnamefont {V.~A.}\ \bibnamefont
  {Kostelecky}}\ and\ \bibinfo {author} {\bibfnamefont {S.}~\bibnamefont
  {Samuel}},\ }\href {\doibase 10.1103/PhysRevD.40.1886} {\bibfield  {journal}
  {\bibinfo  {journal} {Phys. Rev. D}\ }\textbf {\bibinfo {volume} {40}},\
  \bibinfo {pages} {1886} (\bibinfo {year} {1989}{\natexlab{b}})}\BibitemShut
  {NoStop}%
\bibitem [{\citenamefont {Wald}(1980)}]{PhysRevD.21.2742}%
  \BibitemOpen
  \bibfield  {author} {\bibinfo {author} {\bibfnamefont {R.~M.}\ \bibnamefont
  {Wald}},\ }\href {\doibase 10.1103/PhysRevD.21.2742} {\bibfield  {journal}
  {\bibinfo  {journal} {Phys. Rev. D}\ }\textbf {\bibinfo {volume} {21}},\
  \bibinfo {pages} {2742} (\bibinfo {year} {1980})}\BibitemShut {NoStop}%
\end{thebibliography}%

\end{document}